\documentclass[a4paper]{report}
\usepackage[utf8]{inputenc}
\usepackage[T1]{fontenc}
\usepackage{amsmath,amssymb,array}
\usepackage{booktabs}
\usepackage{RJournal}

\begin{document}

\sectionhead{Contributed research article}
\volume{XX}
\volnumber{YY}
\year{20ZZ}
\month{AAAA}

\begin{article}

\title{fasano.franceschini.test: An Implementation of a Multidimensional KS Test in R}
\author{by Connor Puritz, Elan Ness-Cohn, Rosemary Braun}

\maketitle
\abstract{%
The Kolmogorov--Smirnov (KS) test is a nonparametric statistical test used to test for differences between univariate probability distributions. The versatility of the KS test has made it a cornerstone of statistical analysis across many scientific disciplines. However, the test proposed by Kolmogorov and Smirnov does not easily extend to multidimensional distributions. Here we present the \textbf{fasano.franceschini.test} package, an R implementation of a multidimensional two-sample KS test described by \citet{ff1987}. The \textbf{fasano.franceschini.test} package provides a test that is computationally efficient, applicable to data of any dimension and type (continuous, discrete, or mixed), and that performs competitively with similar R packages.
}


\section{Introduction}
\label{introduction}
The Kolmogorov--Smirnov (KS) test is a nonparametric, univariate statistical test designed to assess whether a sample of data is consistent with a given probability distribution (or, in the two-sample case, whether the two samples came from the same underlying distribution). First described by Kolmogorov and Smirnov in a series of papers \citep{Kolmogorov1933, Kolmogorov1933a, Smirnov1936, Smirnov1937, Smirnov1939, Smirnov1944, Smirnov1948}, the KS test is a popular goodness-of-fit test that has found use across a wide variety of scientific disciplines \citep[e.g.][]{atasoy_2017,chiang_2018,hahne_2018,wong_2020,kaczanowska_2021}.

Due to its popularity, several multivariate extensions of the KS test have been described in literature. \citet{justel_1997} proposed a multivariate test based on Rosenblatt's transformation, which reduces to the KS test in the univariate case. While the test statistic is distribution-free, it is difficult to compute in more than two dimensions, and an approximate test with reduced power must be used instead. Furthermore, the test is only applicable in the one-sample case. \citet{heuchenne_2022} proposed to use the Hilbert space-filling curve to define an ordering in $\mathbb{R}^{2}$. The preimage of both samples is computed under the space-filling curve map, and the two-sample KS test is performed on the preimages. While it is theoretically possible to extend this approach to higher dimensions, the authors note that this would be computationally challenging and leave it as an open problem. \citet{naaman_2021} derived a multivariate extension of the DKW inequality and used it to provide estimates of the tail properties of the asymptotic distribution of the KS test statistic in multiple dimensions. While an important theoretical result, practical usage is limited absent a method for computing exact $p$-values.

\citet{peacock_1983} proposed a test which addresses the fact that there are multiple ways to order points in higher dimensions, and thus multiple ways of defining a cumulative distribution function. In one dimension, probability density can be integrated from left to right, resulting in the canonical CDF $P(X<x)$; or from right to left, resulting in the survival function $P(X>x)$. However, since $P(X<x)=1-P(X>x)$ (for continuous random variables), the KS test statistic is independent of this choice. In two dimensions, there are four ways of ordering points, and thus four possible cumulative distribution functions: $P(X<x,Y<y)$, $P(X>x,Y<y)$, $P(X<x,Y>y)$, and $P(X>x,Y>y)$. Since any three are independent, the KS test statistic will depend on which is used. To address this, \citet{peacock_1983} proposed to compute a KS statistic using each possible cumulative distribution function, and to take the test statistic to be the maximum of those.

\citet{peacock_1983} suggested that for a sample $\{(X_{i},Y_{i}):1\leq i\leq n\}$, each KS statistic be maximized over the set of all coordinate-wise combinations $\{(X_{i},Y_{j}):1\leq i,j\leq n\}$. The complexity of computing Peacock's test statistic thus scales cubically with sample size, which is expensive and can become intractable for large sample sizes. \citet{ff1987} proposed a simple change to Peacock's test: instead of maximizing each KS statistic over all coordinate-wise combinations of points in the sample, they are maximized over just the points in the sample itself. This slight change greatly reduces the computational complexity of the test while maintaining a similar power across a variety of alternatives \citep{ff1987,lopes_2007}.

In this article we present the \textbf{fasano.franceschini.test} package, an R implementation of the two-sample Fasano--Franceschini test. Our implementation can be applied to continuous, discrete, or mixed datasets of any size and of any dimension. We first introduce the test by detailing how the test statistic is computed, how we compute it efficiently, and how we compute $p$-values. We then describe the package structure and provide several basic examples illustrating its usage. We conclude by comparing our package to three other CRAN packages implementing multivariate two-sample goodness-of-fit tests.


\section{Fasano--Franceschini test}
\subsection{Two-sample test statistic}
Let
\begin{equation*}
S_{1}=\left\{\mathbf{X}_{1},\dots,\mathbf{X}_{n_{1}}\right\},\;\;\;S_{2}=\left\{\mathbf{Y}_{1},\dots,\mathbf{Y}_{n_{2}}\right\}
\end{equation*}
be samples of i.i.d.\ $d$-dimensional random vectors drawn from unknown distributions $F_{1}$ and $F_{2}$, respectively. The two-sample Fasano--Franceschini test evaluates the null hypothesis
\begin{equation*}
H_{0}:F_{1}=F_{2}
\end{equation*}
against the alternative
\begin{equation*}
H_{1}:F_{1}\neq F_{2}.
\end{equation*}
In their original paper, \citet{ff1987} only considered two- and three-dimensional random vectors, although their test naturally extends to arbitrary dimensions as follows.

For a given point $\mathbf{x}\in\mathbb{R}^{d}$, we define the $i$th open orthant with origin $\mathbf{x}$ as
\begin{equation*}
\mathcal{O}_{i}(\mathbf{x})=\left\{\mathbf{x}'\in\mathbb{R}^{d}\,\rvert\,\mathbf{e}_{ij}(\mathbf{x}_{j}-\mathbf{x}'_{j})>0,\;j=1,\dots,d\right\}
\end{equation*}
where $\mathbf{e}_{i}\in\{-1,1\}^{d}$ is a length $d$ combination of $\pm 1$. For example, in two dimensions, the four combinations $\mathbf{e}_{1}=(1,1)$, $\mathbf{e}_{2}=(-1,1)$, $\mathbf{e}_{3}=(-1,-1)$, and $\mathbf{e}_{4}=(1,-1)$ correspond to quadrants one through four in the plane, respectively. In general there are $2^{d}$ such combinations, corresponding to the $2^{d}$ orthants that divide $\mathbb{R}^{d}$. Using the indicator function
\begin{equation*}
I_{j}(\mathbf{x}\,\rvert\,\mathbf{y})=
\begin{cases}
1,\hfill&\mathbf{x}\in\mathcal{O}_{j}(\mathbf{y})\\
0,\hfill&\mathbf{x}\notin\mathcal{O}_{j}(\mathbf{y})
\end{cases}
\end{equation*}
we define the distance
\begin{equation}
\label{eq:diff}
D(\mathbf{p})=\max_{1\leq j\leq 2^{d}}\left|\frac{1}{n_{1}}\sum_{k=1}^{n_{1}}I_{j}\left(\mathbf{X}_{k}\,\rvert\,\mathbf{p}\right)-\frac{1}{n_{2}}\sum_{k=1}^{n_{2}}I_{j}\left(\mathbf{Y}_{k}\,\rvert\,\mathbf{p}\right)\right|.
\end{equation}
This is similar to the distance used in the two-sample KS test, but takes into account all possible ways of ordering points in $\mathbb{R}^{d}$. Note that this distance does not depend on the enumeration of the orthants. The distance is then maximized over each sample separately, leading to the difference statistics
\begin{equation*}
D_{1}=\max_{1\leq i\leq n_{1}}D(\mathbf{X}_{i})
\end{equation*}
and
\begin{equation*}
D_{2}=\max_{1\leq i\leq n_{2}}D(\mathbf{Y}_{i}).
\end{equation*}
The two-sample Fasano--Franceschini test statistic is then defined as the average of the difference statistics scaled by the sample sizes:
\begin{equation}
\label{eq:ff_stat}
\mathcal{D}=\sqrt{\frac{n_{1}n_{2}}{n_{1}+n_{2}}}\left(\frac{D_{1}+D_{2}}{2}\right).
\end{equation}

\begin{figure}[htbp]
\centering
\includegraphics[scale=0.248]{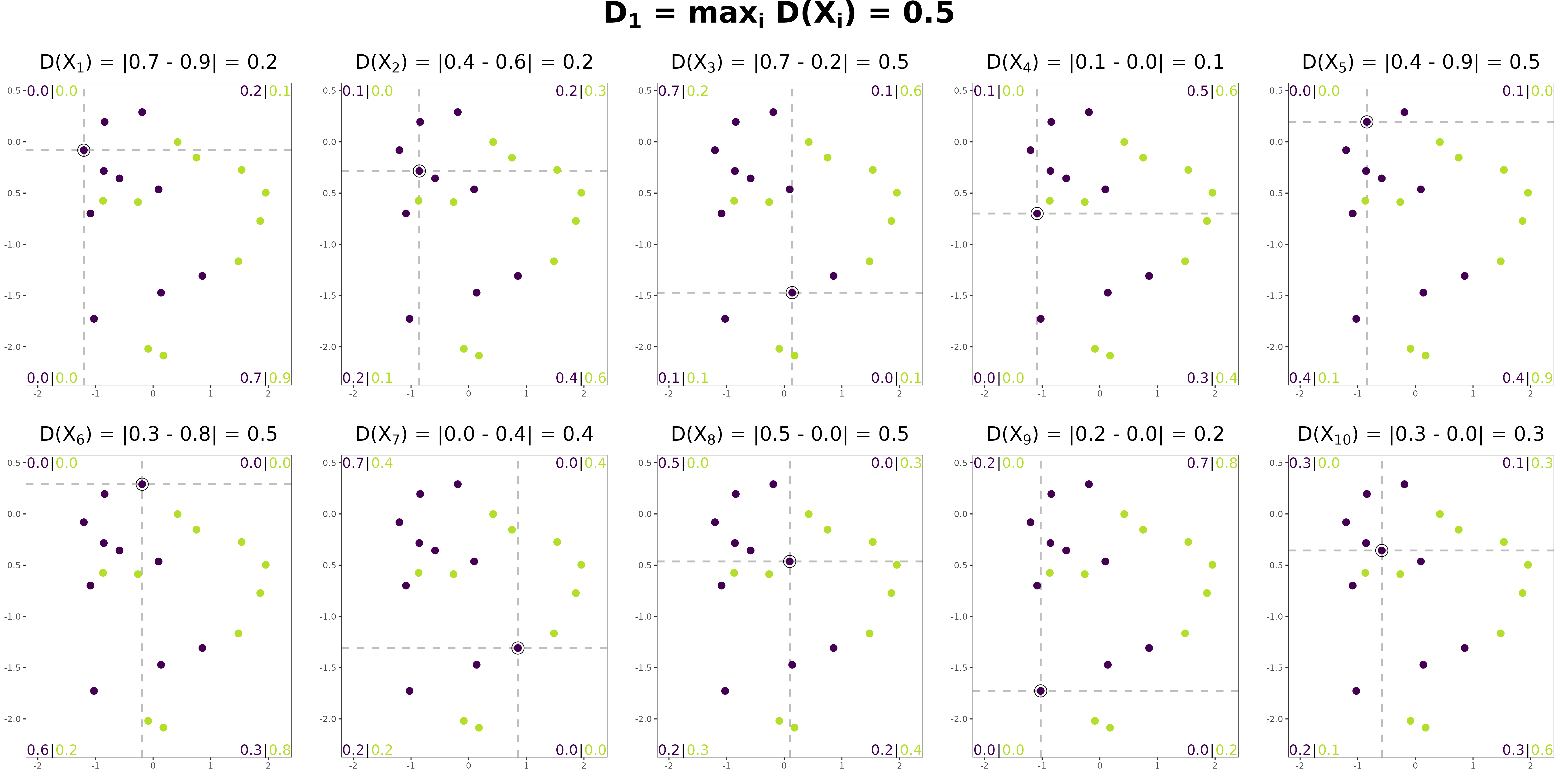}
\caption{Illustration of the computation of the difference statistic $D_{1}$ in two dimensions. Each point in the first sample is used to divide the plane into four quadrants, and both samples are cumulated in each of the four quadrants. The fraction of each sample in each quadrant is shown in the corresponding plot corner, and the maximum difference over all four quadrants is shown above each plot. $D_{1}$ is taken as the maximum of these differences. To compute the Fasano--Franceschini test statistic, the same procedure would need to be repeated, but using points in the second sample to divide the plane instead.}
\label{figure:example_stat}
\end{figure}


\subsection{Computational complexity}
The bulk of the time required to compute the two-sample Fasano--Franceschini test statistic (Eq.~\ref{eq:ff_stat}) is spent evaluating sums of the form
\begin{equation*}
\sum_{\mathbf{x}\in S}I_{j}\left(\mathbf{x}\,\rvert\,\mathbf{y}\right),
\end{equation*}
which count the number of points in a set $S$ that lie in a given $d$-dimensional region. The simplest approach to computing such sums is brute force, where every point $\mathbf{x}\in S$ is checked independently. The orthant a point lies in can be determined using $d$ binary checks, resulting in a time complexity of $O(N^{2})$ (where $N=\max(n_{1},n_{2})$) to evaluate Eq.~(\ref{eq:ff_stat}) for fixed $d$.

Alternatively, we can consider each sum as a single query rather than a sequence of independent ones. Specifically, both sums in Eq.~(\ref{eq:diff}) are orthogonal range counting queries, which ask how many points in a set $S\subset\mathbb{R}^{d}$ lie in an axis-aligned box $(x_{1},x_{1}')\times\dots \times(x_{d},x_{d}')$. Range counting is an important problem in the field of computational geometry, and as such a variety of data structures have been described to provide efficient solutions \citep{geometry_2008}. One solution, first introduced by \citet{bentley_decomposable_1979}, is a multi-layer binary search tree termed a range tree. Other slightly more efficient data structures have been proposed for range counting, but range trees are well suited for our purposes, particularly because their construction scales easily to arbitrary dimensions \citep{bentley_decomposable_1979, geometry_2008}.

A range tree can be constructed on a set of $n$ points in $d$-dimensional space using $O(n\log^{d-1}n)$ space in $O(n\log^{d-1}n)$ time. The number of points that lie in an axis-aligned box can be reported in $O(\log^{d}n)$ time, and this time can be further reduced to $O(\log^{d-1}n)$ (when $d>1$) using fractional cascading \citep{geometry_2008}. To compute the two-sample Fasano--Franceschini test statistic, we construct one range tree for each of the two samples, and then query each tree $2^{d}$ times. Thus the total time complexity to compute the test statistic using range trees for fixed $d$ is $O(N\log^{d-1}N)$, where again $N=\max(n_{1},n_{2})$.

\begin{figure}[htbp]
\centering
\includegraphics[scale=0.50]{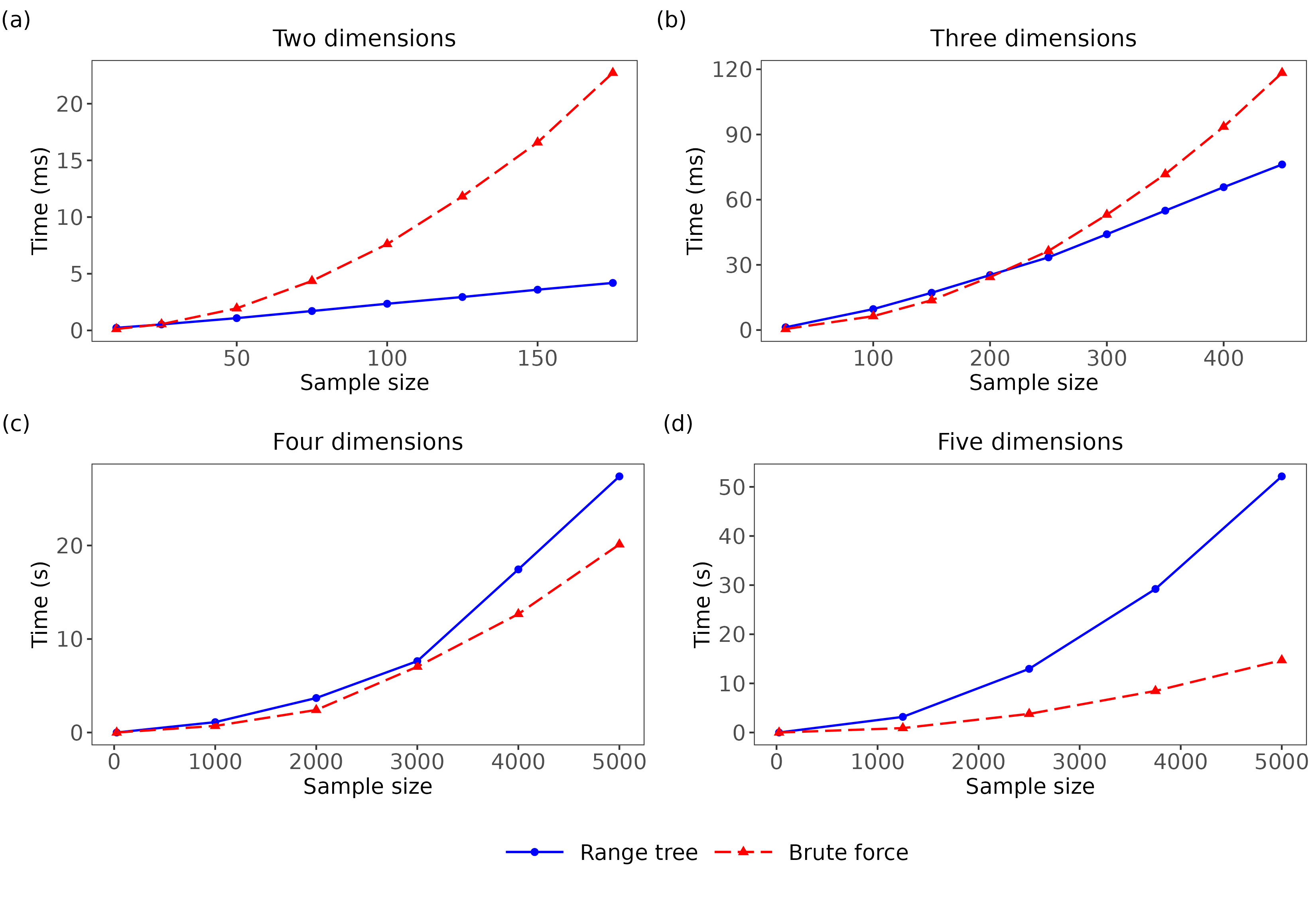}
\caption{Time to compute the Fasano--Franceschini test statistic as a function of sample size, comparing the brute force and range tree methods for data of dimensions two through five. Points represent the mean time of $200$ evaluations. In each plot, samples are taken to be the same size and are drawn from multivariate standard normal distributions.}
\label{figure:stat_time}
\end{figure}

As the range tree method has a better asymptotic time complexity than the brute force method, we expect it to perform better for larger sample sizes. However, for smaller sample sizes, the cost of building the range trees can outweigh the benefit gained by more efficient querying. For each dimension, we sought to determine the sample size $N^{*}$ at which the range tree method becomes more efficient than the brute force method (Figure~\ref{figure:stat_time}). For $d=2$, $N^{*}\approx 25$; for $d=3$, $N^{*}\approx 200$; for $d=4,5$, and presumably all higher dimensions, $N^{*}>5000$. As goodness-of-fit tests are generally applied to samples of much smaller sizes than this, we stopped benchmarking here.

Based on these benchmarking results, our package automatically selects which of the two methods is likely faster based on the dimension and samples sizes of the supplied data. However, as we used equal sample sizes during benchmarking, and since computation time can vary depending on the geometry of the samples, the selected method may not actually be fastest. If users are interested in performing benchmarking for their specific dataset, the argument \code{nPermute} can be set equal to $0$, which bypasses the permutation test and only computes the test statistic.


\subsection{Significance testing}
To the best of our knowledge, no results have been published concerning the distribution of the Fasano--Franceschini test statistic. Any analysis would likely be complicated by the fact that, unlike the KS test statistic, the Fasano--Franceschini test statistic is not distribution free \citep{ff1987}. In their original paper, \citet{ff1987} did not attempt any analytical analysis and instead performed Monte Carlo simulations to estimate critical values of their test statistic for various two- and three-dimensional distributions. By fitting a curve to their results, \citet{press2007} proposed an explicit formula for $p$-values in the two-dimensional case. However, this formula is only approximate, and its accuracy degrades as sample sizes decrease or the true $p$-value becomes large (greater than $0.2$). While this will still allow a simple rejection decision at any common significance level, it is sometimes useful to quantify large $p$-values more exactly (such as if one was to do a cross-study concordance analysis comparing $p$-values between studies as in \citet{ness-cohn_2020}). Effort could be made to improve this approximation, however it is still only valid in two dimensions, and thus an alternative method would be needed in higher dimensions.

To allow the \textbf{fasano.franceschini.test} package to be applicable to as broad a class of problems as possible, we compute $p$-values using a permutation test. Under the null hypothesis, the two samples were drawn from the same underlying distribution, and a permutation test leverages this to compute the null distribution of the test statistic. Permutation tests are distribution free, and can be applied to continuous, discrete, or mixed data of any dimension. The test procedure is as follows:
\begin{enumerate}
\item Compute the test statistic $\mathcal{D}$ for the original samples $S_{1}$ and $S_{2}$.
\item Pool the two samples, and label each element according to which sample it belongs to.
\item Permute the labels, and split the pooled sample into two new samples $S_{1}^{i}$ and $S_{2}^{i}$ according to the new labels.
\item Compute the test statistic $\mathcal{D}_{i}$ for $S_{1}^{i}$ and $S_{2}^{i}$.
\item Repeat steps (3-4) for every permutation of the labels.
\item The $p$-value is fraction of test statistics $\mathcal{D}_{i}$ at least as large as $\mathcal{D}$.
\end{enumerate}

However, as the sample sizes increase to even modest values, the total number of permutations of the labels increases rapidly, and it quickly becomes computationally infeasible to compute the test statistic for every permutation. Thus instead of considering all permutations, we select a fixed number of permutations $M$ with replacement and compute a Monte Carlo approximation of the $p$-value, given by
\begin{equation*}
\hat{p}=\frac{1+\sum_{i=1}^{M}I(\mathcal{D}_{i}\geq \mathcal{D})}{1+M}
\end{equation*}
where
\begin{equation*}
I(x\geq y)=\begin{cases} 1,& x\geq y\\ 0,& x< y. \end{cases}
\end{equation*}
If permutations are selected without replacement, this estimator is exact. However, if permutations are selected with replacement, this estimator is slightly more conservative than the exact estimator \citep{phipson_2010}. Unless sample sizes are small, the loss of power will be minimal as the likelihood of selecting the same permutation multiple times will be negligible.

We select permutations with replacement primarily to circumvent the computationally expensive step of ensuring that repeated permutations are not selected. An additional benefit is that we are easily able compute a confidence interval for the true permutation $p$-value, as the number of test statistics for permuted samples at least as large as $\mathcal{D}$ is distributed binomially with a probability of success equal to the true permutation test $p$-value \citep{good2005}. We compute the confidence interval using the \texttt{binom.test} function from the \textbf{stats} package, which computes an exact binomial confidence interval as given in \citet{clopper_pearson_1934}.


\section{Package overview}
The \textbf{fasano.franceschini.test} package is written primarily in C++, and interfaces with R using \textbf{Rcpp} \citep{rcpp}. The permutation test is parallelized using \textbf{RcppParallel} \citep{rcppparallel}. The package consists of one function, \code{fasano.franceschini.test}, for performing the two-sample Fasano--Franceschini test. The arguments of this function are described below.
\begin{itemize}
\item \code{S1} and \code{S2}: the two samples to compare. Both should be either numeric \code{matrix} or \code{data.frame} objects with the same number of columns.
\item \code{nPermute}: the number of permuted samples to generate when estimating the permutation test $p$-value. The default is $100$. If set equal to $0$, the permutation test is bypassed and only the test statistic is computed.
\item \code{threads}: the number of threads to use when performing the permutation test. The default is one thread. This parameter can also be set to \code{"auto"}, which uses the value returned by \code{RcppParallel::defaultNumThreads()}.
\item \code{seed}: an optional seed for the pseudorandom number generator (PRNG) used during the permutation test.
\item \code{p.conf.level}: the confidence level for the confidence interval of the permutation test $p$-value. The default is $0.95$.
\item \code{verbose}: whether to display a progress bar while performing the permutation test. The default is \code{TRUE}. This functionality is only available when \code{threads = 1}.
\item \code{method}: an optional \code{character} indicating which method to use to compute the test statistic. The two methods are \code{'r'} (range tree) and
\code{'b'} (brute force). Both methods return the same results but may vary in computation speed. If this argument is not passed, the sample sizes and dimension
of the data are used to infer which method is likely faster.
\end{itemize}
The output is an object of the class \code{htest}, and consists of the following components:
\begin{itemize}
\item \code{statistic}: the value of the test statistic $\mathcal{D}$.
\item \code{estimate}: the value of the difference statistics $D_{1}$ and $D_{2}$.
\item \code{p.value}: a Monte-Carlo approximation of the permutation test $p$-value.
\item \code{conf.int}: a binomial confidence interval for the permutation test $p$-value.
\item \code{method}: the name of the test (i.e. \code{'Fasano-Francheschini Test'}).
\item \code{data.name}: the names of the original data objects.
\end{itemize}


\section{Examples}
Here we demonstrate the basic usage and features of the \textbf{fasano.franceschini.test} package. We begin by loading the necessary libraries and setting a seed for reproducibility.
\begin{example}
> library(fasano.franceschini.test)
> library(MASS)
> set.seed(1)	
\end{example}
Note that to produce reproducible results, we need to set two seeds: the \texttt{set.seed} function sets the seed in R, ensuring we draw reproducible samples; and the seed passed as an argument to the \texttt{fasano.franceschini.test} function sets the seed for the C++ PRNG, ensuring we compute reproducible $p$-value estimates.

As a first example, we draw two samples from a bivariate standard normal distribution. The Fasano--Franceschini test fails to reject the null hypothesis --- that the samples were drawn from the same distribution --- at an $\alpha=0.05$ significance level.

\begin{example}
> S1 <- mvrnorm(n = 100, mu = c(0,0), Sigma = diag(2))
> S2 <- mvrnorm(n = 150, mu = c(0,0), Sigma = diag(2))
> fasano.franceschini.test(S1, S2, seed = 2, verbose = FALSE)

	Fasano-Francheschini Test

data:  S1 and S2
D = 0.85206, p-value = 0.8416
95 percent confidence interval:
 0.7555271 0.9066534
sample estimates:
  D1   D2 
0.11 0.11 
\end{example}

We next draw two samples from bivariate normal distributions with identical covariance matrices but different locations. The Fasano--Franceschini test rejects the null hypothesis at an $\alpha=0.05$ significance level.

\begin{example}
> S3 <- mvrnorm(n = 225, mu = c(0,0), Sigma = diag(2))
> S4 <- mvrnorm(n = 152, mu = c(0.2,0.2), Sigma = diag(2))
> fasano.franceschini.test(S3, S4, seed = 3, verbose = FALSE)

	Fasano-Francheschini Test

data:  S3 and S4
D = 2.0212, p-value = 0.009901
95 percent confidence interval:
 0.00025064 0.05393235
sample estimates:
       D1        D2 
0.2109649 0.2134503 
\end{example}

However, we note that $\alpha=0.05$ is contained in the $p$-value confidence interval. To be careful, we rerun the test with $200$ permutations instead of the default $100$, in which case both the $p$-value estimate and the right endpoint of its confidence interval are strictly less than $\alpha=0.05$.

\begin{example}
> fasano.franceschini.test(S3, S4, nPermute = 200, seed = 3, verbose = FALSE)

	Fasano-Francheschini Test

data:  S3 and S4
D = 2.0212, p-value = 0.004975
95 percent confidence interval:
 0.0001259513 0.0274064298
sample estimates:
       D1        D2 
0.2109649 0.2134503 
\end{example}


\section{Comparison with other R packages}
In this section, we compare the \textbf{fasano.franceschini.test} package with three other CRAN packages that perform multivariate two-sample goodness-of-fit tests.

\subsection{Peacock.test}
The \textbf{Peacock.test} package \citep{peacockR} provides functions to compute Peacock's test statistic \citep{peacock_1983} in two and three dimensions. As no function is provided to compute $p$-values, we cannot directly compare the performance of this package with the \textbf{fasano.franceschini.test} package. However, a thorough treatment of the power of both Peacock and Fasano--Franceschini tests can be found in both the primary literature \citep{peacock_1983,ff1987} and in a subsequent benchmarking paper \citep{lopes_2007}, which found that the two tests have similar power across a variety of alternatives.

\subsection{cramer}
The \textbf{cramer} package \citep{cramerR} implements the two-sample test described in \citet{baringhaus_2004}, which the authors refer to as the Cram\'er test. The Cram\'er test statistic is based on the Euclidean inter-point distances between the two samples, and is given by
\begin{equation*}
T_{m,n}=\frac{mn}{m+n}\left(\frac{2}{mn}\sum_{i=1}^{m}\sum_{j=1}^{n}\phi\left(\left\lVert\mathbf{X}_{i}-\mathbf{Y}_{j}\right\rVert_{2}^{2}\right)-\frac{1}{m^{2}}\sum_{i,j=1}^{m}\phi\left(\left\lVert\mathbf{X}_{i}-\mathbf{X}_{j}\right\rVert_{2}^{2}\right)-\frac{1}{n^{2}}\sum_{i,j=1}^{m}\phi\left(\left\lVert\mathbf{Y}_{i}-\mathbf{Y}_{j}\right\rVert_{2}^{2}\right)\right)
\end{equation*}
for samples $\{\mathbf{X}_{1},\dots,\mathbf{X}_{m}\}$ and $\{\mathbf{Y}_{1},\dots,\mathbf{Y}_{n}\}$. In the documentation, several options for the function $\phi$ are given, with the default being
\begin{equation*}
\phi(x)=\sqrt{x}/2.
\end{equation*}
This statistic is not distribution-free, and several methods are provided to compute $p$-values. By default, $p$-values are estimated using a bootstrapping procedure.

\subsection{diproperm}
The \textbf{diproperm} package \citep{dipropermR} implements the DiProPerm test introduced by \citet{diproperm_original}. The test first trains a binary linear classifier to determine a separating hyperplane between the two samples. The data are then projected onto the normal vector to the hyperplane, and the test statistic is taken to be a univariate statistic of the projected data (by default the absolute difference of means). Like in the \textbf{fasano.franceschini.test} package, significance is determined using a permutation test.

\subsection{Power comparison}
To compare the \textbf{fasano.franceschini.test} package with the \textbf{cramer} and \textbf{diproperm} packages, we performed power analyses using three classes of alternatives: location alternatives, where the means of the marginals are varied; dispersion alternatives, where the variances of the marginals are varied; and copula alternatives, where the marginals remain fixed but the copula joining them is varied.

For location and dispersion alternatives, we used multivariate normal distributions. We denote the $d$-dimensional normal distribution with mean $\boldsymbol\mu\in\mathbb{R}^{d}$ and covariance matrix $\boldsymbol\Sigma\in\mathbb{R}^{d\times d}$ by $N_{d}(\boldsymbol\mu,\boldsymbol\Sigma)$, and sample from it using the \textbf{MASS} package \citep{mass}. The $d\times d$ identity matrix, which is sometimes used as a covariance matrix, is denoted as $\mathbf{I}_{d}$. For copula alternatives, we consider the Gaussian copula with correlation matrix
\begin{equation*}
[P(\rho)]_{ij}=\begin{cases}
\rho,&i\neq j\\
1,&i=j
\end{cases}
\end{equation*}
and the Clayton copula with parameter $\theta\in[-1,\infty)\setminus\{0\}$. We denote the $d$-dimensional distribution with standard normal marginals joined by a Gaussian copula with correlation matrix $P(\rho)$ by $G_{d}(\rho)$. We denote the $d$-dimensional distribution with standard normal marginals joined by a Clayton copula with parameter $\theta$ by $C_{d}(\theta)$. Both distributions are sampled from using the \textbf{copula} package \citep{copulaR}. For all power analyses performed, power was approximated using $1000$ replications, a significance level of $\alpha=0.05$ was used, all samples were of size $50$, and all R functions implementing tests were called using their default arguments.

\begin{figure}[htbp]
\centering
\includegraphics[scale=0.50]{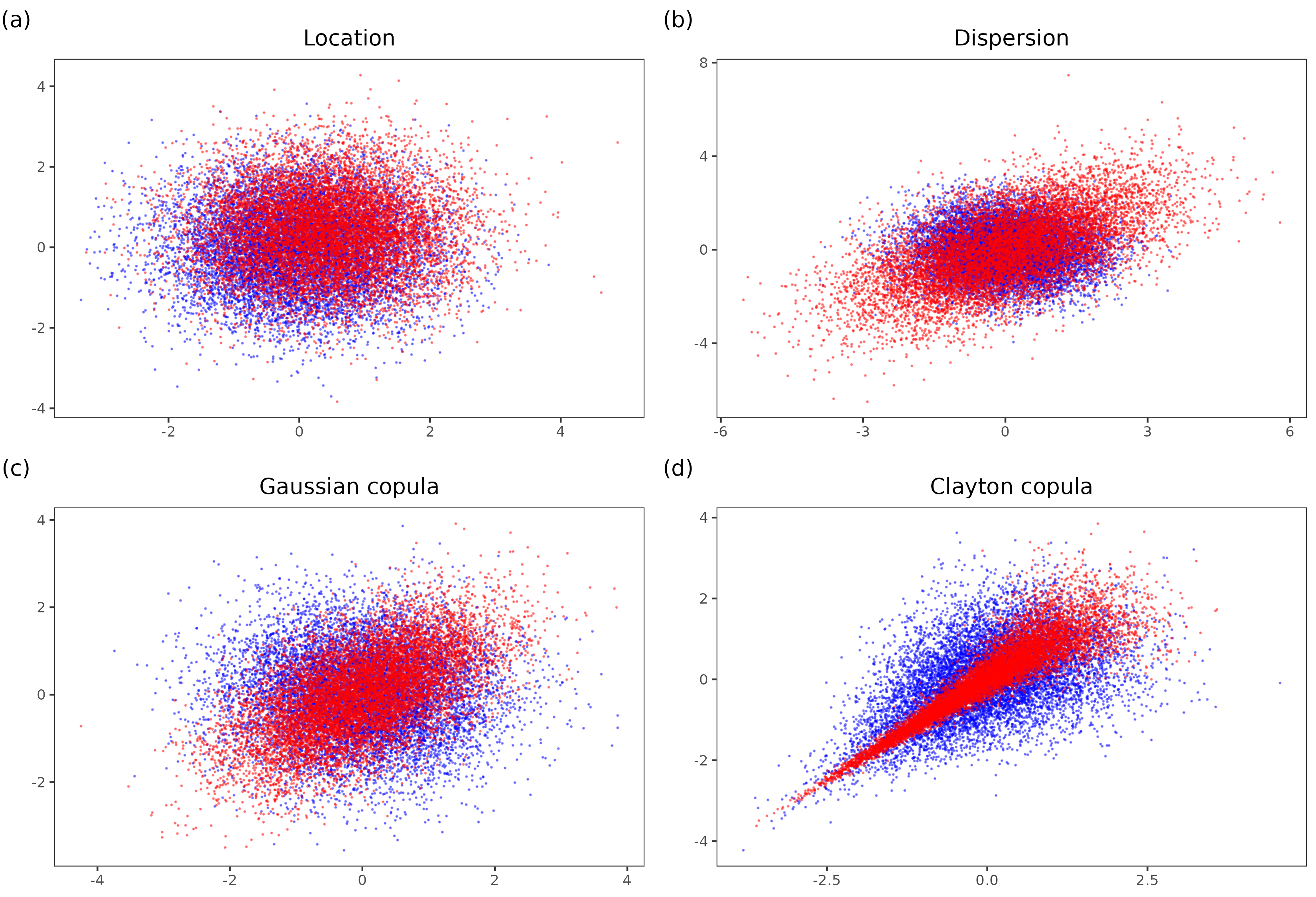}
\caption{Visualization of the distributions used in power analyses. Each plot shows two samples consisting of $10000$ points each. The first sample $S_{1}$ is shown in blue, and the second sample $S_{2}$ is shown in red. (a) $S_{1}\sim N_{2}(\mathbf{0},\mathbf{I}_{2})$ and $S_{2}\sim N_{2}(\mathbf{0.4},\mathbf{I}_{2})$. (b) $S_{1}\sim N_{2}(\mathbf{0},\mathbf{I}_{2})$ and $S_{2}\sim N_{2}(\mathbf{0},\mathbf{I}_{2}+1.5)$. (c) $S_{1}\sim G_{2}(0)$ and $S_{2}\sim G_{d}(0.6)$. (d) $S_{1}\sim C_{2}(1)$ and $S_{2}\sim C_{2}(8)$.}
\label{figure:power_dist}
\end{figure}

\begin{figure}[htbp]
\centering
\includegraphics[scale=0.50]{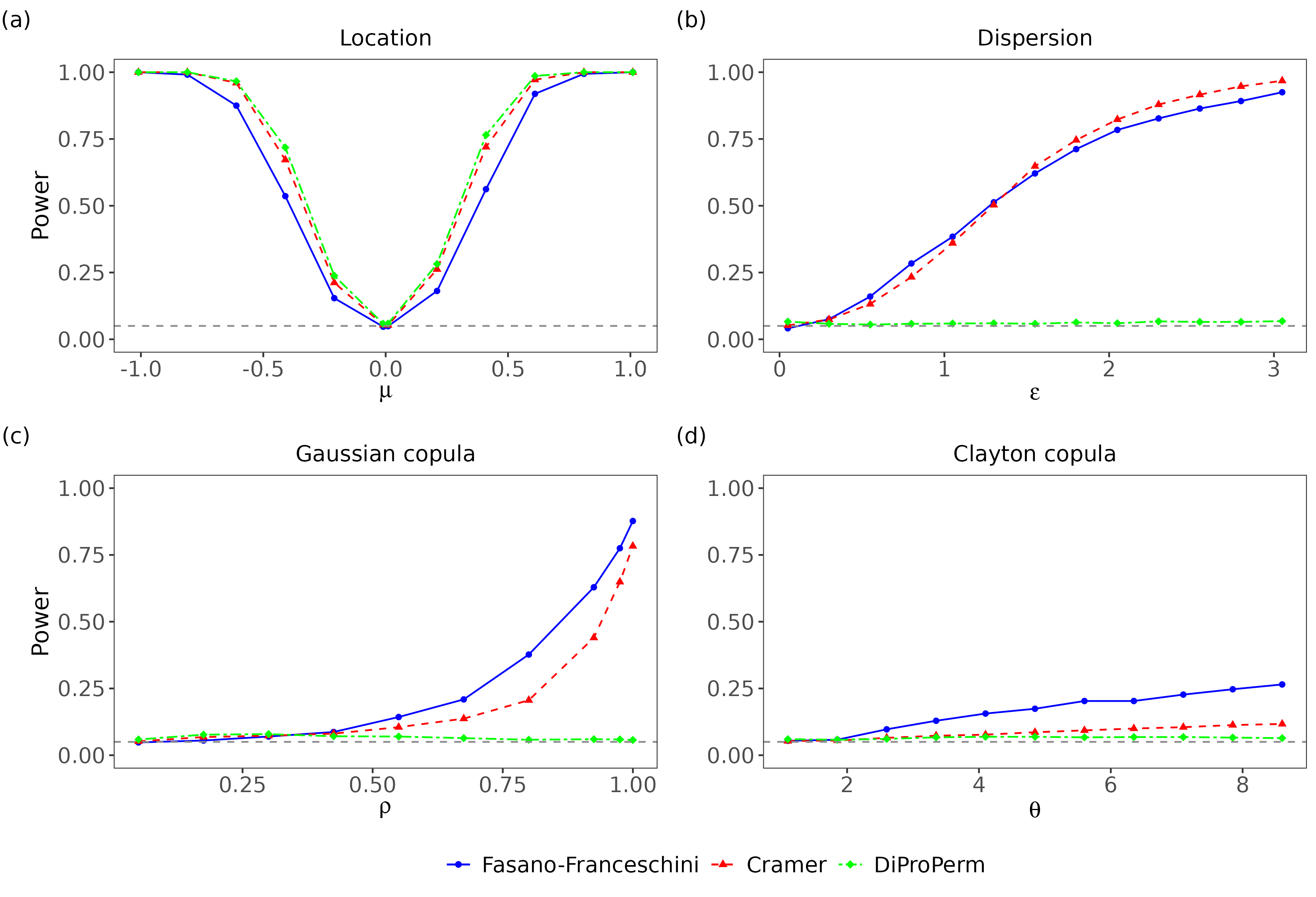}
\caption{Comparison of power of the Fasano--Franceschini, Cram\'er, and DiProPerm tests on various bivariate alternatives. (a) Location alternatives, with $S_{1}\sim N_{2}(\mathbf{0},\mathbf{I}_{2})$ and $S_{2}\sim N_{2}(\boldsymbol\mu,\mathbf{I}_{2})$. (b) Dispersion alternatives, with $S_{1}\sim N_{2}(\mathbf{0},\mathbf{I}_{2})$ and $S_{2}\sim N_{2}(\mathbf{0},\mathbf{I}_{2}+\varepsilon)$. (c) Gaussian copula alternatives, with $S_{1}\sim G_{2}(0)$ and $S_{2}\sim G_{2}(\rho)$. (d) Clayton copula alternatives, with $S_{1}\sim C_{2}(1)$ and $S_{2}\sim C_{2}(\theta)$.}
\label{figure:bivariate_power}
\end{figure}

We first examined the power of the tests on various bivariate alternatives. All three tests had similar power across location alternatives, although the Cram\'er and DiProPerm tests did tend to slightly outperform the Fasano--Franceschini test. Across dispersion alternatives, the Cram\'er and Fasano--Franceschini tests had very similar powers. On both copula alternatives, the Fasano--Franceschini test had a consistently higher power than the Cram\'er test. The DiProPerm test was unable to achieve a power above the significance level of $\alpha=0.05$ on any of the dispersion or copula alternatives. This is likely due to the fact that in these instances, there is significant overlap between the high density regions of the two distributions, making it difficult to find a separating hyperplane between samples drawn from them.

\begin{figure}[htbp]
\centering
\includegraphics[scale=0.50]{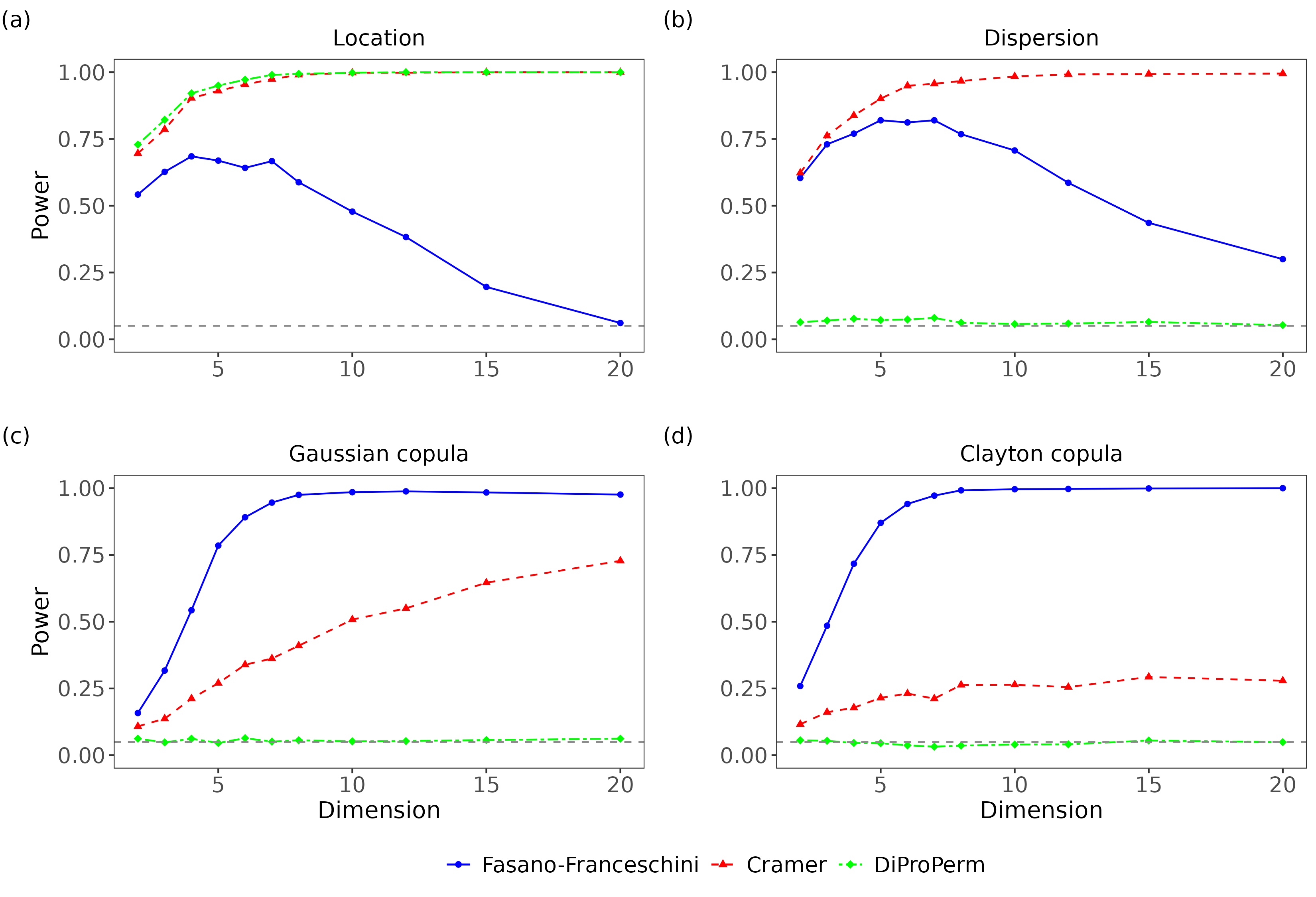}
\caption{Comparison of power of the Fasano--Franceschini, Cram\'er, and DiProPerm tests on fixed alternatives as the dimension of the data increases. (a) Location alternative, with $S_{1}\sim N_{d}(\mathbf{0},\mathbf{I}_{d})$ and $S_{2}\sim N_{d}(\mathbf{0.4},\mathbf{I}_{d})$. (b) Dispersion alternative, with $S_{1}\sim N_{d}(\mathbf{0},\mathbf{I}_{d})$ and $S_{2}\sim N_{d}(\mathbf{0},\mathbf{I}_{d}+1.5)$. (c) Gaussian copula alternative, with $S_{1}\sim G_{d}(0)$ and $S_{2}\sim G_{d}(0.6)$. (d) Clayton copula alternative, with $S_{1}\sim C_{d}(1)$ and $S_{2}\sim C_{d}(8)$.}
\label{figure:dimension_power}
\end{figure}

We next examined how the power of the three tests varied when the two sampling distributions were kept fixed but the dimension of the data increased. On the location alternative, the power of the Cram\'er and DiProPerm tests was quite similar, monotonically increasing to one as dimension increased. The power of the Fasano--Franceschini increased until $d=5$ and then monotonically decreased to $\alpha=0.05$ by $d=20$. We see similar results for the Cram\'er and Fasano--Franceschini tests on the dispersion alternative. On copula alternatives, both the Cram\'er and Fasano--Franceschini tests have monotonically increasing power as dimension is increased. However, whereas the Fasano--Franceschini test is able to achieve a power of nearly one near $d=10$ on both alternatives, the Cram\'er test's power grows at a much slower rate. The DiProPerm test is still unable to attain a power above $\alpha=0.05$ on the dispersion alternatives or either of the copula alternatives.

Overall, the Cram\'er and DiProPerm tests perform better than the Fasano--Franceschini test on location alternatives, especially as dimension increases. On dispersion alternatives, the Fasano--Franceschini and Cram\'er tests have comparable performance for low dimensions, but the Cram\'er test maintains a higher power for high dimensions. However, in these cases the marginal distributions differ, and thus a multivariate test is not strictly necessary as univariate tests could be applied to the marginals independently (with a multiple testing correction) to detect the difference between the multivariate distributions. On copula alternatives, where a multivariate test is required, the Fasano--Franceschini test consistently outperformed both the Cram\'er and DiProPerm tests.


\section{Summary}
This paper introduces the \textbf{fasano.franceschini.test} package, an R implementation of the multidimensional two-sample goodness-of-fit test described by \citet{ff1987}. We provide users with a computationally efficient test that is applicable to data of any dimension and of any type (continuous, discrete, or mixed), and that demonstrates competitive performance with similar R packages. Complete package documentation and source code are available via the Comprehensive R Archive Network (CRAN) at \url{https://cran.r-project.org/web/packages/fasano.franceschini.test} and the package website at \url{https://nesscoder.github.io/fasano.franceschini.test}.


\section{Computational details}
The results in this paper were obtained using R 4.1.1 with the packages \textbf{fasano.franceschini.test} 2.1.1, \textbf{diproperm} 0.2.0, \textbf{cramer} 0.9-3, \textbf{MASS} 7.3-54, and \textbf{copula} 1.1-0. All computations were done using the Quest high performance computing facility at Northwestern University. R itself and all package dependencies are available from CRAN at \url{https://cran.r-project.org}.


\section{Acknowledgments}
This research was supported in part through the computational resources and staff contributions provided for the Quest high performance computing facility at Northwestern University which is jointly supported by the Office of the Provost, the Office for Research, and Northwestern University Information Technology.

Funding for this work was provided by the Simons Foundation/SFARI (597491-RWC01), the National Science Foundation (1764421-01), and the National Institute of Health/National Institute of Aging (1R01AG068579-01).


\bibliography{fasano-franceschini-bib}

\begin{thebibliography}{34}
\providecommand{\natexlab}[1]{#1}
\providecommand{\url}[1]{\texttt{#1}}
\expandafter\ifx\csname urlstyle\endcsname\relax
  \providecommand{\doi}[1]{doi: #1}\else
  \providecommand{\doi}{doi: \begingroup \urlstyle{rm}\Url}\fi

\bibitem[Allaire et~al.(2022)Allaire, Francois, Ushey, Vandenbrouck, Geelnard,
  and {Intel}]{rcppparallel}
J.~Allaire, R.~Francois, K.~Ushey, G.~Vandenbrouck, M.~Geelnard, and {Intel}.
\newblock \emph{RcppParallel: Parallel Programming Tools for 'Rcpp'}, 2022.
\newblock URL \url{https://CRAN.R-project.org/package=RcppParallel}.
\newblock R package version 5.1.5.

\bibitem[Allmon et~al.(2021)Allmon, Marron, and Hudgens]{dipropermR}
A.~G. Allmon, J.~Marron, and M.~G. Hudgens.
\newblock \emph{diproperm: Conduct Direction-Projection-Permutation Tests and
  Display Plots}, 2021.
\newblock URL \url{https://CRAN.R-project.org/package=diproperm}.
\newblock R package version 0.2.0.

\bibitem[Atasoy et~al.(2017)Atasoy, Roseman, Kaelen, Kringelbach, Deco, and
  Carhart-Harris]{atasoy_2017}
S.~Atasoy, L.~Roseman, M.~Kaelen, M.~L. Kringelbach, G.~Deco, and R.~L.
  Carhart-Harris.
\newblock Connectome-harmonic decomposition of human brain activity reveals
  dynamical repertoire re-organization under {LSD}.
\newblock \emph{Scientific Reports}, 7\penalty0 (1):\penalty0 1--18, 2017.
\newblock URL \url{https://doi.org/10.1038/s41598-017-17546-0}.

\bibitem[Baringhaus and Franz(2004)]{baringhaus_2004}
L.~Baringhaus and C.~Franz.
\newblock On a new multivariate two-sample test.
\newblock \emph{Journal of Multivariate Analysis}, 88\penalty0 (1):\penalty0
  190--206, 2004.
\newblock ISSN 0047-259X.
\newblock URL \url{https://doi.org/10.1016/S0047-259X(03)00079-4}.

\bibitem[Bentley(1979)]{bentley_decomposable_1979}
J.~L. Bentley.
\newblock Decomposable searching problems.
\newblock \emph{Information Processing Letters}, 8\penalty0 (5):\penalty0
  244--251, 1979.
\newblock ISSN 0020-0190.
\newblock URL \url{https://doi.org/10.1016/0020-0190(79)90117-0}.

\bibitem[Chiang et~al.(2018)Chiang, Mazdiyasni, and AghaKouchak]{chiang_2018}
F.~Chiang, O.~Mazdiyasni, and A.~AghaKouchak.
\newblock Amplified warming of droughts in southern united states in
  observations and model simulations.
\newblock \emph{Science Advances}, 4\penalty0 (8):\penalty0 eaat2380, 2018.
\newblock URL \url{https://doi.org/10.1126/sciadv.aat2380}.

\bibitem[Clopper and Pearson(1934)]{clopper_pearson_1934}
C.~J. Clopper and E.~S. Pearson.
\newblock {The Use of Confidence or Fiducial Limits Illustrated in the Case of
  the Binomial}.
\newblock \emph{Biometrika}, 26\penalty0 (4):\penalty0 404--413, 1934.
\newblock ISSN 00063444.
\newblock URL \url{https://doi.org/10.2307/2331986}.

\bibitem[de~Berg et~al.(2008)de~Berg, Cheong, van Kreveld, and
  Overmars]{geometry_2008}
M.~de~Berg, O.~Cheong, M.~van Kreveld, and M.~Overmars.
\newblock \emph{Computational {Geometry}: {Algorithms} and {Applications}}.
\newblock Springer Berlin Heidelberg, 3rd edition, 2008.
\newblock ISBN 978-3-540-77974-2.
\newblock URL \url{https://doi.org/10.1007/978-3-540-77974-2}.

\bibitem[Eddelbuettel et~al.(2022)Eddelbuettel, Francois, Allaire, Ushey, Kou,
  Russell, Ucar, Bates, and Chambers]{rcpp}
D.~Eddelbuettel, R.~Francois, J.~Allaire, K.~Ushey, Q.~Kou, N.~Russell,
  I.~Ucar, D.~Bates, and J.~Chambers.
\newblock \emph{Rcpp: Seamless R and C++ Integration}, 2022.
\newblock URL \url{https://CRAN.R-project.org/package=Rcpp}.
\newblock R package version 1.0.9.

\bibitem[Fasano and Franceschini(1987)]{ff1987}
G.~Fasano and A.~Franceschini.
\newblock A multidimensional version of the {Kolmogorov}-{Smirnov} test.
\newblock \emph{Monthly Notices of the Royal Astronomical Society},
  225\penalty0 (1):\penalty0 155--170, 03 1987.
\newblock ISSN 0035-8711.
\newblock URL \url{https://doi.org/10.1093/mnras/225.1.155}.

\bibitem[Franz(2019)]{cramerR}
C.~Franz.
\newblock \emph{cramer: Multivariate Nonparametric Cramer-Test for the
  Two-Sample-Problem}, 2019.
\newblock URL \url{https://CRAN.R-project.org/package=cramer}.
\newblock R package version 0.9-3.

\bibitem[Good(2005)]{good2005}
P.~I. Good.
\newblock \emph{{Permutation, Parametric and Bootstrap Tests of Hypotheses}}.
\newblock Springer-Verlag New York, 2005.
\newblock ISBN 978-0-387-27158-3.
\newblock URL \url{https://doi.org/10.1007/b138696}.

\bibitem[Hahne et~al.(2018)Hahne, Schweisfurth, Koppe, and Farina]{hahne_2018}
J.~M. Hahne, M.~A. Schweisfurth, M.~Koppe, and D.~Farina.
\newblock Simultaneous control of multiple functions of bionic hand prostheses:
  Performance and robustness in end users.
\newblock \emph{Science Robotics}, 3\penalty0 (19):\penalty0 eaat3630, 2018.
\newblock URL \url{https://doi.org/10.1126/scirobotics.aat3630}.

\bibitem[Heuchenne and Mordant(2022)]{heuchenne_2022}
C.~Heuchenne and G.~Mordant.
\newblock Using space filling curves to compare two multivariate distributions
  with distribution-free tests.
\newblock \emph{Journal of Computational and Applied Mathematics},
  416:\penalty0 114494, Dec. 2022.
\newblock ISSN 0377-0427.
\newblock URL \url{https://doi.org/10.1016/j.cam.2022.114494}.

\bibitem[Hofert et~al.(2022)Hofert, Kojadinovic, Maechler, and Yan]{copulaR}
M.~Hofert, I.~Kojadinovic, M.~Maechler, and J.~Yan.
\newblock \emph{copula: Multivariate Dependence with Copulas}, 2022.
\newblock URL \url{https://CRAN.R-project.org/package=copula}.
\newblock R package version 1.1-0.

\bibitem[Justel et~al.(1997)Justel, Peña, and Zamar]{justel_1997}
A.~Justel, D.~Peña, and R.~Zamar.
\newblock A multivariate {Kolmogorov}-{Smirnov} test of goodness of fit.
\newblock \emph{Statistics \& Probability Letters}, 35\penalty0 (3):\penalty0
  251--259, 1997.
\newblock ISSN 0167-7152.
\newblock URL \url{https://doi.org/10.1016/S0167-7152(97)00020-5}.

\bibitem[Kaczanowska et~al.(2021)Kaczanowska, Beury, Gopalan, Tycko, Qin,
  Clements, Drake, Nwanze, Murgai, Rae, Ju, Alexander, Kline, Contreras,
  Wessel, Patel, Hannenhalli, Kelly, and Kaplan]{kaczanowska_2021}
S.~Kaczanowska, D.~W. Beury, V.~Gopalan, A.~K. Tycko, H.~Qin, M.~E. Clements,
  J.~Drake, C.~Nwanze, M.~Murgai, Z.~Rae, W.~Ju, K.~A. Alexander, J.~Kline,
  C.~F. Contreras, K.~M. Wessel, S.~Patel, S.~Hannenhalli, M.~C. Kelly, and
  R.~N. Kaplan.
\newblock Genetically engineered myeloid cells rebalance the core immune
  suppression program in metastasis.
\newblock \emph{Cell}, 184\penalty0 (8):\penalty0 2033--2052.e21, 2021.
\newblock ISSN 0092-8674.
\newblock URL \url{https://doi.org/10.1016/j.cell.2021.02.048}.

\bibitem[Kolmogorov(1933{\natexlab{a}})]{Kolmogorov1933}
A.~N. Kolmogorov.
\newblock {Sulla Determinazione Empirica di Una Legge di Distribuzione}.
\newblock \emph{Giornale dell'Istituto Italiano degli Attuari}, 4:\penalty0
  83--91, 1933{\natexlab{a}}.

\bibitem[Kolmogorov(1933{\natexlab{b}})]{Kolmogorov1933a}
A.~N. Kolmogorov.
\newblock {\"Uber die Grenzwerts\"atze der Wahrscheinlichkeitsrechnung}.
\newblock \emph{{Bull. Acad. Sci. URSS}}, 3:\penalty0 363--372,
  1933{\natexlab{b}}.
\newblock URL \url{http://mi.mathnet.ru/eng/izv5009}.

\bibitem[Lopes et~al.(2007)Lopes, Reid, and Hobson]{lopes_2007}
R.~H.~C. Lopes, I.~Reid, and P.~R. Hobson.
\newblock The two-dimensional {Kolmogorov}-{Smirnov} test.
\newblock In \emph{XI International Workshop on Advanced Computing and Analysis
  Techniques in Physics Research}, 2007.
\newblock URL \url{https://bura.brunel.ac.uk/handle/2438/1166}.

\bibitem[Naaman(2021)]{naaman_2021}
M.~Naaman.
\newblock On the tight constant in the multivariate
  {Dvoretzky}–{Kiefer}–{Wolfowitz} inequality.
\newblock \emph{Statistics \& Probability Letters}, 173:\penalty0 109088, 2021.
\newblock ISSN 0167-7152.
\newblock URL \url{https://doi.org/10.1016/j.spl.2021.109088}.

\bibitem[Ness-Cohn et~al.(2020)Ness-Cohn, Iwanaszko, Kath, Allada, and
  Braun]{ness-cohn_2020}
E.~Ness-Cohn, M.~Iwanaszko, W.~L. Kath, R.~Allada, and R.~Braun.
\newblock {TimeTrial: An Interactive Application for Optimizing the Design and
  Analysis of Transcriptomic Time-Series Data in Circadian Biology Research}.
\newblock \emph{Journal of Biological Rhythms}, 35\penalty0 (5):\penalty0
  439--451, 2020.
\newblock URL \url{https://doi.org/10.1177/0748730420934672}.
\newblock PMID: 32613882.

\bibitem[Peacock(1983)]{peacock_1983}
J.~A. Peacock.
\newblock Two-dimensional goodness-of-fit testing in astronomy.
\newblock \emph{Monthly Notices of the Royal Astronomical Society},
  202\penalty0 (3):\penalty0 615--627, 03 1983.
\newblock ISSN 0035-8711.
\newblock URL \url{https://doi.org/10.1093/mnras/202.3.615}.

\bibitem[Phipson and Smyth(2010)]{phipson_2010}
B.~Phipson and G.~K. Smyth.
\newblock Permutation p-values should never be zero: calculating exact p-values
  when permutations are randomly drawn.
\newblock \emph{Statistical Applications in Genetics and Molecular Biology},
  9\penalty0 (1), 2010.
\newblock ISSN 1544-6115.
\newblock URL \url{https://doi.org/10.2202/1544-6115.1585}.

\bibitem[Press et~al.(2007)Press, Teukolsky, Vetterling, and
  Flannery]{press2007}
W.~H. Press, S.~A. Teukolsky, W.~T. Vetterling, and B.~P. Flannery.
\newblock \emph{Numerical Recipes: The Art of Scientific Computing}.
\newblock Cambridge University Press, USA, 3rd edition, 2007.
\newblock ISBN 0521880688.
\newblock URL \url{https://doi.org/10.1145/1874391.187410}.

\bibitem[Ripley(2021)]{mass}
B.~Ripley.
\newblock \emph{MASS: Support Functions and Datasets for Venables and Ripley's
  MASS}, 2021.
\newblock URL \url{https://CRAN.R-project.org/package=MASS}.
\newblock R package version 7.3-54.

\bibitem[Smirnov(1936)]{Smirnov1936}
N.~V. Smirnov.
\newblock {Sur la distribution de $\omega^2$ (criterium de M.R. v. Mises)}.
\newblock \emph{Com. Rend. Acad. Sci. (Paris)}, 202:\penalty0 449--452, 1936.

\bibitem[Smirnov(1937)]{Smirnov1937}
N.~V. Smirnov.
\newblock On the distribution of the mises $\omega^2$ criterion [in {Russian}].
\newblock \emph{{Rec. Math. N.S. [Mat. Sbornik]}}, 2:\penalty0 973--993, 1937.
\newblock URL \url{http://mi.mathnet.ru/eng/msb/v44/i5/p973}.

\bibitem[Smirnov(1939)]{Smirnov1939}
N.~V. Smirnov.
\newblock On the deviations of the empirical distribution curve [in {Russian}].
\newblock \emph{{Rec. Math. N.S. [Mat. Sbornik]}}, 6\penalty0 (48):\penalty0
  3--26, 1939.
\newblock URL \url{http://mi.mathnet.ru/eng/msb/v48/i1/p3}.

\bibitem[Smirnov(1944)]{Smirnov1944}
N.~V. Smirnov.
\newblock Approximate laws of distribution of random variables from empirical
  data.
\newblock \emph{{Uspehi Matem. Nauk}}, 10:\penalty0 179--206, 1944.

\bibitem[Smirnov(1948)]{Smirnov1948}
N.~V. Smirnov.
\newblock Table for estimating the goodness of fit of empirical distributions.
\newblock \emph{The Annals of Mathematical Statistics}, 1948.
\newblock ISSN 0003-4851.
\newblock URL \url{https://doi.org/10.1214/aoms/1177730256}.

\bibitem[Wei et~al.(2016)Wei, Lee, Wichers, and Marron]{diproperm_original}
S.~Wei, C.~Lee, L.~Wichers, and J.~S. Marron.
\newblock {Direction-Projection-Permutation for High-Dimensional Hypothesis
  Tests}.
\newblock \emph{Journal of Computational and Graphical Statistics}, 25\penalty0
  (2):\penalty0 549--569, 2016.
\newblock URL \url{https://doi.org/10.1080/10618600.2015.1027773}.

\bibitem[Wong and Collins(2020)]{wong_2020}
F.~Wong and J.~J. Collins.
\newblock Evidence that coronavirus superspreading is fat-tailed.
\newblock \emph{Proceedings of the National Academy of Sciences}, 117\penalty0
  (47):\penalty0 29416--29418, 2020.
\newblock URL \url{https://doi.org/10.1073/pnas.2018490117}.

\bibitem[Xiao(2016)]{peacockR}
Y.~Xiao.
\newblock \emph{Peacock.test: Two and Three Dimensional Kolmogorov-Smirnov
  Two-Sample Tests}, 2016.
\newblock URL \url{https://CRAN.R-project.org/package=Peacock.test}.
\newblock R package version 1.0.

\end{thebibliography}


\address{Connor Puritz\\
  Department of Engineering Sciences and Applied Mathematics\\
  Northwestern University\\
  Evanston, IL 60208\\
  ORCiD: 0000-0001-7602-0444\\
  Email: \email{connorpuritz2025@u.northwestern.edu}}

\address{Elan Ness-Cohn\\
  Department of Molecular Biosciences\\
  and\\
  NSF-Simons Center for Quantitative Biology\\
  Northwestern University\\
  Evanston, IL 60208\\
  ORCiD: 0000-0002-3935-6667\\
  Email: \email{elan.ness-cohn@northwestern.edu}\\
  Website: \email{https://www.nesscoder.com}}

\address{Rosemary Braun\\
  Department of Molecular Biosciences,\\
  Department of Engineering Sciences and Applied Mathematics,\\
  Department of Physics and Astronomy,\\
  Northwestern Institute of Complex Systems,\\
  and\\
  NSF-Simons Center for Quantitative Biology\\
  Northwestern University\\
  Evanston, IL 60208\\
  ORCiD: 0000-0001-9668-9866\\
  Email: \email{rbraun@northwestern.edu}\\
  Website: \email{https://sites.northwestern.edu/braunlab/}}

\end{article}

\end{document}